\input{aipcheck}

\documentclass[
    ,final            
  ]
  {aipproc}

\layoutstyle{8x11single}

\begin{document}

\title{Scalar and Pseudoscalar Glueballs Revisited}


\author{Hai-Yang Cheng}{
  address={Institute of Physics, Academia Sinica, Taipei, Taiwan 115, Republic of China}
  ,altaddress={and Physics Department, Brookhaven National Laboratory, Upton, NY 11973} 
}

\begin{abstract}
Using two simple and robust inputs to constrain the mixing matrix of the isosinglet scalar mesons $f_0(1710)$, $f_0(1500)$, $f_0(1370)$, we have shown that in the SU(3) symmetry limit,
$f_0(1500)$ becomes a pure SU(3) octet and is degenerate with
$a_0(1450)$, while $f_0(1370)$ is mainly an SU(3) singlet with a
slight mixing with the scalar glueball which is the primary
component of $f_0(1710)$. These features remain essentially
unchanged even when SU(3) breaking is taken into account.
We have deduced the mass of the pseudoscalar glueball $G$ from an
$\eta$-$\eta'$-$G$ mixing formalism based on the anomalous Ward
identity for transition matrix elements. With the inputs from the
recent KLOE experiment, we found a solution for the pseudoscalar
glueball mass around $(1.4\pm 0.1)$ GeV. This affirms that $\eta(1405)$, having a large production rate in the radiative $J/\psi$ decay and not seen in $\gamma\gamma$ reactions, is indeed a leading candidate for the pseudoscalar glueball. It is much lower than the results from quenched lattice QCD ($>2.0$ GeV).
\end{abstract}

\maketitle


\section{Introduction}\label{aba:sec1}
The existence of glueballs is a unique prediction of QCD as a
confining theory.  The latest lattice calculation of the glueball spectroscopy shows that the lightest glueballs are scalar, tensor and pseudoscalar glueballs with masses of order 1710, 2390 and 2560 MeV, respectively \cite{Chen}. Since the lattice calculation was done in the quenched approximation, the predicted masses are for pure glueballs in the Yang-Mills gauge theory. The question is what happens to the glueballs in the presence of quark degrees of freedom ? First, a glueball will mix with the ordinary meson with the same quantum numbers so that a pure glueball does not exist in nature. Since the glueball is hidden somewhere in the quark sector, this is one of main reasons why it is so elusive. Second, we shall show that the mass of the pseudoscalar glueball could be drastically affected by the dynamic fermion effect.

\section{Scalar glueball}
It is generally believed that the scalar glueball is hidden somewhere in the isosinglet scalar mesons with masses above 1 GeV. The argument goes as follows. Many scalar mesons with masses lower than 2 GeV have been observed and they can be classified into two nonets: one nonet with mass below or close to 1 GeV, such as $\sigma,~\kappa$, $f_0(980)$ and $a_0(980)$ that are generally believed to be composed mainly of four quarks and the other nonet with mass above 1 GeV such as $K_0^*(1430)$, $a_0(1450)$ and two isosinglet scalar mesons. This means that not all three isosinglet scalars $f_0(1710)$, $f_0(1500)$, $f_0(1370)$ can be accommodated in the $q\bar q$ nonet picture. One of them should be primarily a scalar glueball.\footnote{It has been suggested that $f_0(600)$ or the $\sigma$ state is a good candidate for the scalar glueball (see e.g. \cite{Ochs}).
Since a pure glueball state (scalar or pseudoscalar) cannot decay into a photon pair or a massless quark pair to the leading order,
the identification of $f_0(600)$ with a scalar glueball seems to be unlikely in view of its broad width, of order $600\sim 1000$ MeV, and its sizable partial width to $\gamma\gamma$ of order 4keV \cite{PDG}. In the conventional tetraquark picture of $f_0(600)$, its broad width can be naturally understood as fall-apart decays.}

Among the three isoscalar mesons, it has been quite controversial as to which of these
is the dominant scalar glueball. It was suggested that
$f_0(1500)$ is primarily a scalar glueball in \cite{Close1}, due
partly to the fact that $f_0(1500)$, discovered in $p\bar{p}$
annihilation at LEAR, has decays to $\eta\eta$ and $\eta\eta'$
which are relatively large compared to that of
$\pi\pi$ and that the earlier quenched lattice
calculations predict the scalar glueball mass to be
around $1550$ MeV~\cite{bsh93}. Furthermore, because of the small production of
$\pi\pi$ in $f_0(1710)$ decay compared to that of $K\bar K$, it
has been thought that $f_0(1710)$ is primarily $s\bar s$
dominated. In contrast, the smaller production rate of $K\bar K$
relative to $\pi\pi$ in $f_0(1370)$ decay leads to the conjecture
that $f_0(1370)$ is governed by the non-strange light quark
content. Based on the above observations, a flavor-mixing scheme was proposed \cite{Close1} to consider the glueball and $q\bar q$
mixing in the neutral scalar mesons $f_0(1710)$, $f_0(1500)$ and
$f_0(1370)$.  Best $\chi^2$ fits to the measured scalar meson
masses and their branching fractions of strong decays have been
performed in several references by Amsler, Close and
Kirk~\cite{Close1}, Close and Zhao~\cite{Close2}, and He {\it et
al.}~\cite{He}. A typical mixing matrix in this scheme is
\cite{Close2}
\begin{eqnarray} \label{eq:Close}
 \left(\begin{matrix} {f_0(1370) \cr f_0(1500) \cr f_0(1710) \cr } \end{matrix}\right)=
\left( \begin{matrix}{ -0.91 & -0.07 & 0.40 \cr
                 -0.41 & 0.35 & -0.84 \cr
                0.09 & 0.93 & 0.36 \cr}
                  \end{matrix}\right)\left(\begin{matrix}{|N\rangle \cr
 |S\rangle \cr |G\rangle \cr} \end{matrix}\right).
\end{eqnarray}
A common feature of these analyses is that, before mixing, the
$s\bar{s}$ quarkonium mass $M_S$ is larger than the glueball mass $M_G$
which, in turn, is larger than the
$N(\equiv(u\bar{u}+d\bar{d})/\sqrt{2})$ mass $M_N$, with $M_G$
close to 1500 MeV and $M_S-M_N$ of the order of $200\sim 300$ MeV.

However, the above scenario encounters several difficulties: (i) The isovector scalar meson $a_0(1450)$ is now
confirmed to be the $q\bar{q}$ meson in the lattice
calculation~\cite{Mathur}. As such, the degeneracy of $a_0(1450)$
and $K_0^*(1430)$, which has a strange quark, cannot be explained
if $M_S$ is larger than $M_N$ by $\sim 250$ MeV. (ii) The most
recent quenched lattice calculation with improved action and
lattice spacings extrapolated to the continuum favors a larger
scalar glueball mass close to 1700 MeV~\cite{Chen}. (iii) If
$f_0(1710)$ is dominated by the $s\bar s$ content, the decay
$J/\psi\to \phi f_0(1710)$ is expected to have a rate larger than
that of $J/\psi\to \omega f_0(1710)$. Experimentally, it is other
way around: the rate for $\omega f_0(1710)$ production is about 3
times that of $J/\psi\to \phi f_0(1710)$. (iv)  If $f_0(1500)$ has the largest scalar glueball component, one expects  the
$\Gamma(J/\psi\to \gamma f_0(1500))$ decay rate to be
substantially larger than that of $\Gamma(J/\psi\to \gamma
f_0(1710))$. Again, experimentally, the opposite is true.

In our work \cite{CCL}, we employed two simple and robust
results as the input for the mass matrix which is essentially the
starting point for the mixing model between scalar mesons and the
glueball. First of all, we know empirically that flavor SU(3) is an approximate symmetry in the scalar meson sector above 1 GeV. The near degeneracy of $K_0^*(1430)$, $a_0(1470)$, and $f_0(1500)$ has been observed. In the scalar charmed meson sector, $D_{s0}^*(2317)$ and $D_0^*(2308)$ have similar masses even though the former contains a strange quark. It is most likely that the same phenomenon also holds in the scalar bottom meson sector.  This unusual behavior is not understood
as far as we know and it serves as a challenge to the existing
hadronic models.
Second, an improved quenched lattice calculation
of the glueball spectrum at the infinite volume and continuum
limits based on much larger and finer lattices have been carried
out~\cite{Chen}. The mass of the scalar glueball is calculated to
be $m(0^{++})=1710\pm50\pm 80$ MeV.   This suggests that $M_G$
should be close to 1700 MeV rather than 1500 MeV from the earlier
lattice calculations~\cite{bsh93}.


We begin by considering exact SU(3) symmetry as a first
approximation for the mass matrix.  In this case, two of the mass eigenstates are identified
with $a_0(1450)$ and $f_0(1500)$ which are degenerate with the
mass $M$. Taking $M$ to be the experimental mass of $1474\pm 19$
MeV \cite{PDG}, it is a good approximation for the mass of
$f_0(1500)$ at $1507\pm 5$ MeV \cite{PDG}. Thus, in the limit of
exact SU(3) symmetry, $f_0(1500)$ is an SU(3) isosinglet octet
state and is degenerate with $a_0(1450)$. In the absence of glueball-quarkonium mixing, $f_0(1370)$
becomes a pure SU(3) singlet  and $f_0(1710)$
the pure glueball.  When the
glueball-quarkonium mixing is turned on, there will be some
mixing between the glueball and the SU(3)-singlet $q\bar{q}$ . The mass
shift of $f_0(1370)$ and $f_0(1710)$ due to mixing is only of order 10 MeV.
Since the SU(3) breaking effect is expected to be
weak, it can be treated perturbatively.

\subsection{Chiral suppression}
If $f_0(1710)$ is primarily a pseudoscalar glueball, it is naively expected that $\Gamma(G\to\pi\pi)/\Gamma(G\to K\bar K)\approx 0.9$ after phase space correction due to the flavor independent coupling of $G$ to $PP$. However, experimentally there is a relatively large suppression of $\pi\pi$ production
relative to $K\bar K$ in $f_0(1710)$ decay, $R(f_0(1710))\equiv\Gamma(f_0(1710)\to \pi \pi)/\Gamma(f_0(1710)\to K\bar K)=0.41^{+0.11}_{-0.17}$ \cite{BES} or even smaller.
To explain the large disparity between $\pi\pi$ and $K\bar K$ production in scalar glueball decays, it was noticed long time ago by Carlson et al. \cite{Carlson}, by Cornwall and Soni \cite{Cornwall} and revitalized recently by Chanowitz \cite{Chanowitz} that a pure scalar or pseudoscalar glueball cannot decay into a quark-antiquark pair in the chiral limit, i.e., $A(G\to q\bar q)\propto m_q$.  Since the current strange quark mass is an order of magnitude larger than $m_u$ and $m_d$, decay to $K\bar K$ is largely favored over $\pi\pi$.  However, chiral suppression for the ratio $\Gamma(G\to \pi\pi)/\Gamma(G\to K\bar K)$ at the hadron level should not be so strong as the current quark mass ratio $m_u/m_s$.  It has been suggested \cite{Chao} that $m_q$ should be interpreted as the scale of chiral symmetry breaking.

Whether or not $G\to\pi\pi$ is subject to chiral suppression is a controversial issue because of the hadronization process from $G\to q\bar q$ to $G\to\pi\pi$ and the possible competing $G\to q\bar qq\bar q$ mechanism \cite{Carlson,Chao,Jin,Chanowitz:reply}. The only reliable method for tackling with the nonperturbative effects is lattice QCD. An earlier lattice calculation \cite{Sexton} did support the chiral-suppression effect with the result $\Gamma(G\to \eta\eta)>\Gamma(G\to K\bar K)\gg \Gamma(G\to\pi\pi)$. Although the errors are large, the lattice result did show a sizable deviation from the flavor-symmetry limit. If the chiral suppression of scalar-glueball decay is confirmed by (quenched or unquenched) lattice calculations, the experimental measurement of $R(f_0(1500))=4.1\pm0.4$ alone \cite{PDG} will be sufficient to rule out the possibility of $f_0(1500)$ being a glueball as the ratio $R(f_0(1500))$
should be less than 3/4 due to chiral suppression.  Likewise, the identification of $f_0(600)$ with a scalar glueball is also very unlikely owing to its broad width.

Guided by the lattice calculations for chiral suppression in $G\to
PP$ \cite{Sexton}, we have performed a best $\chi^2$ fit to the measured masses and branching fractions. The mixing matrix obtained in our model has
the form:
\begin{eqnarray}  \label{eq:wf}
 \left(\begin{matrix} { f_0(1370) \cr f_0(1500) \cr f_0(1710) \cr} \end{matrix}\right)=
\left( \begin{matrix} { 0.78 & 0.51 & -0.36 \cr
                 -0.54 & 0.84 & 0.03 \cr
                0.32 & 0.18 & 0.93 \cr}
                  \end{matrix}\right)\left(\begin{matrix}{|N\rangle \cr
 |S\rangle \cr |G\rangle \cr}\end{matrix}\right).
\end{eqnarray}
It is evident that $f_0(1710)$ is composed primarily of the scalar
glueball, $f_0(1500)$ is close to an SU(3) octet, and $f_0(1370)$
consists of an approximate SU(3) singlet with some glueball
component ($\sim 10\%$). Unlike $f_0(1370)$, the glueball content
of $f_0(1500)$ is very tiny because an SU(3) octet does not mix
with the scalar glueball. Because the $n\bar
n$ content is more copious than $s\bar s$ in $f_0(1710)$, it is
natural that $J/\psi\to\omega f_0(1710)$ has a rate larger than
$J/\psi\to \phi f_0(1710)$. Our prediction of $\Gamma(J/\psi\to
\omega f_0(1710))/\Gamma(J/\psi\to \phi f_0(1710))=4.1$ is
consistent with the observed value of $3.3\pm1.3$ \cite{PDG}. Moreover,  if $f_0(1710)$ is composed mainly
of the scalar glueball, it will be expected that
$\Gamma(J/\psi\to \gamma f_0(1710))\gg \Gamma(J/\psi\to \gamma
f_0(1500))$, a relation borne out by experiment.
It is interesting to compare the mixing matrices (\ref{eq:Close}) and (\ref{eq:wf}). In the model of Close and Zhao \cite{Close2}, $f_0(1710)$ is dominated by the $s\bar s$ quarkonium in order to explain the suppression of $R(f_0(1710))$.
 In our scheme, $f_0(1710)$ has the smallest content of $s\bar s$ and the smallness of $R(f_0(1710))$ arises from the chiral suppression of scalar glueball decay. Although $f_0(1500)$ in our model has the largest content of $s\bar s$, the $K\bar K$ production is largely suppressed relative to $\pi\pi$,
 $R(f_0(1500))=3[(\alpha/( \alpha+\beta)]^2(p_\pi/ p_K)$,
where $f_0(1500)=\alpha(|u\bar u\rangle+|d\bar d\rangle)+\beta|s\bar s\rangle$ and $p_h$ is the c.m. momentum of the hadorn $h$. In SU(3) limit, $\beta=-2\alpha$ and this leads to $R(f_0(1500))=3.9$\,, in agreement with experiment.

\section{Pseudoscalar glueball}
In 1980, Mark II observed  a
resonance with a mass around 1440 MeV in the radiative $J/\psi$ decay \cite{iota} which was subsequently named $\iota(1440)$ by Mark II and Crystal Ball Collaborations \cite{CB}.
Shortly after the Mark II experiment, $\iota(1440)$ now known as
$\eta(1405)$
was proposed to be a leading candidate for the pseudoscalar glueball. (For an excellent review of the $E(1420)$ and $\iota(1440)$ mesons,
see \cite{MCU06}.) Indeed $\eta(1405)$ behaves like a glueball in its productions and
decays because it has a large
production rate in the radiative $J/\psi$ decay and is not seen in
$\gamma\gamma$ reactions. Besides $\eta(1405)$, other states
with masses below 2 GeV have also been proposed as the candidates,
such as $\eta(1760)$  and $X(1835)$.

However, the pseudoscalar glueball interpretation
for $\eta(1405)$ is not favored by most of the theoretical calculations. For example,
quenched lattice gauge
calculations predict the mass of the $0^{-+}$ state to be
above 2 GeV in \cite{GSB93} and around 2.6 GeV in
\cite{Morningstar,Chen}. It is not favored by the sum-rule analysis
with predictions higher than 1.8 GeV \cite{sum,GG97} either (for a review of the glueball spectroscopy in various approaches, see \cite{Mathieu}). Hence, we are encountering an embarrassing situation that although experimentally $\eta(1405)$ is a favored candidate for the pseudoscalar glueball, theorists seem to prefer to have a
$0^{-+}$ state heavier than the scalar glueball. The motivation of our recent work \cite{CLL} is to see if we can learn something about the glueball mass by studying the $\eta-\eta'-G$ mixing.

The $\eta-\eta'$ mixing has been well studied by Feldmann, Kroll and Stech \cite{FKS}. We extend the FKS formalism
to include the pseudoscalar glueball $G$. In the FKS scheme, the
conventional singlet-octet basis and the quark-flavor basis have
been proposed. For the latter, the $q\bar q\equiv (u\bar u+d\bar
d)/\sqrt{2}$ and $s\bar s$ flavor states, labeled by the $\eta_q$
and $\eta_s$ mesons, respectively, are defined. The physical states $\eta$, $\eta'$
and $G$ and their decay constants are related to that of the octet, singlet, and unmixed glueball
states $\eta_8$, $\eta_1$ and $g$, respectively, through the
rotation
\begin{eqnarray}\label{qs}
   \left( \begin{array}{c}
    |\eta\rangle \\ |\eta'\rangle\\|G\rangle
   \end{array} \right)
   = U(\phi,\phi_G) \left( \begin{array}{c}
    |\eta_8\rangle \\ |\eta_1\rangle\\|g\rangle
   \end{array} \right) \;, \qquad \left(
\begin{array}{cc}
f_\eta^q & f_\eta^s \\
f_{\eta'}^q & f_{\eta'}^s \\
f_G^q &f_G^s
\end{array} \right)=
U(\phi,\phi_G) \left(
\begin{array}{cc}
f_q & f_q^s \\
f_s^q & f_s \\
f_g^q & f_g^s
\end{array} \right)
\;,\label{fpi}
\end{eqnarray}
where $\phi=\theta+54.7^\circ$ with $\theta$ being the $\eta-\eta'$ mixing angle in the octet-singlet basis, and $\phi_G$ is the mixing angle between $G$ and $\eta_1$; that is, we have assumed that $\eta_8$ does not mix with the glueball. The decay constants for the physical and flavor states are defined by
\begin{eqnarray}
   \langle 0|\bar q\gamma^\mu\gamma_5 q|\eta_q(P)\rangle
   = -\frac{i}{\sqrt2}\,f_q\,P^\mu,    & &\langle 0|\bar q\gamma^\mu\gamma_5 q|\eta_s(P),g(P)\rangle
   = -\frac{i}{\sqrt2}\,f_{s,g}^q\,P^\mu \;, \nonumber\\
   \langle 0|\bar s\gamma^\mu\gamma_5 s|\eta_s(P)\rangle
   = -i f_s\,P^\mu,  & &\langle 0|\bar s\gamma^\mu\gamma_5 s|\eta_q(P),g(P)\rangle
   = -i f_{q,g}^s\,P^\mu \;.\label{deffq}
\end{eqnarray}

Applying the equations of motion for the anomalous Ward identity
\begin{eqnarray}
   \partial_\mu(\bar q\gamma^\mu\gamma_5 q) = 2im_q\,\bar q\gamma_5 q
   +\frac{\alpha_s}{4\pi}\,G_{\mu\nu}\,\widetilde{G}^{\mu\nu},
\end{eqnarray}
between vacuum and $|\eta\rangle$, $|\eta'\rangle$ and $|G\rangle$, we derive six equations for many unknowns. Hence we have to reply on the large $N_c$ counting rules to solve the equations step by step. To the leading order of $1/N_c$ expansion, we found  that the ratio of two of the equations leads to
\begin{eqnarray}
\frac{c\theta (s\phi-\sqrt{2/3}c\theta\Delta_G)m_{\eta'}^2-s\theta
(c\phi+\sqrt{2/3}s\theta\Delta_G)^2m_\eta^2 -\sqrt{2/3} c\phi_G
m_G^2}{c\theta (c\phi-\sqrt{1/3}c\theta\Delta_G)m_{\eta'}^2+s\theta
(s\phi-\sqrt{1/3}s\theta\Delta_G)^2m_\eta^2 -\sqrt{1/3}c\phi_G
m_G^2}=\frac{\sqrt{2}f_s}{f_q},  \label{rg}
\end{eqnarray}
where $\Delta_G=1-\cos\phi_G$ and $c\phi$ ($s\phi$) is the shorthand notation for $\cos\phi$
($\sin\phi$) and similarly for others.  This simple equation tells us that the pseudoscalar glueball mass $m_G$ can be determined provided that the mixing angle $\phi_G$ and the ratio $f_s/f_q$ are known. Note that the $\phi_G$ dependence appears at
order of $\Delta_G\approx \phi_G^2$ for small $\phi_G$. So the solution for $m_G$ is stable against the most
uncertain input $\phi_G$. The mixing angles $\phi$ and $\phi_G$
have been measured by KLOE \cite{KLOE} from the $\phi \to \gamma\eta, \gamma\eta'$ decays. Using the updated results $\phi=(40.4\pm 0.6)^\circ$ and $\phi_G=(20\pm 3)^\circ$
obtained by KLOE \cite{KLOE09} in conjunction with $f_s/f_q=1.352\pm 0.007$, we
derive the pseudoscalar glueball mass from Eq.~(\ref{rg}) to be
\begin{equation} \label{mG}
m_G=(1.4\pm0.1)~{\rm GeV}.
\end{equation}
The proximity of the predicted $m_G$ to the mass of
$\eta(1405)$ and other properties of $\eta(1405)$ make it a very
strong candidate for the pseudoscalar glueball. Using the above-mentioned values for $\phi$ and $\phi_G$, we obtain the $\eta-\eta'-G$ mixing matrix
\begin{eqnarray}
\left( \begin{array}{c}
    |\eta\rangle \\ |\eta'\rangle\\|G\rangle
   \end{array} \right)=\left(\begin{array}{ccc}
  0.749 & -0.657 &0.085\\
  0.600 & 0.728 & 0.331 \\
  -0.279 &-0.197 & 0.940
\end{array} \right)\left( \begin{array}{c}
    |\eta_8\rangle \\ |\eta_1\rangle\\|g\rangle
   \end{array} \right).
\end{eqnarray}

Our next task is to check the stability and robustness of our prediction when higher
order effects in $1/N_c$ are included.
It turns out that the above simple formula Eq. (\ref{rg}) still holds even after keeping the OZI-correcting decay constants, as long as they obey the relations $f_g^q = \sqrt{2} f_g^s$ and $f_s^q =f_q^s$.
Therefore, if excluding the solutions with large and negative
$m_{sg}^2$,
the range $(1.4 \pm 0.1)$ GeV of the pseudoscalar glueball mass
obtained in Eq. (\ref{mG}) will be more or less respected.

One may feel very  uncomfortable with our solution for $m_G$ as both lattice QCD and QCD sum rules indicate a pseudoscalar glueball heavier than the scalar one.
The point is that lattice calculations so far were
performed under the quenched approximation without the fermion
determinants. It is believed that dynamical fermions will have a
significant effect in the pseudoscalar channel, because they raise
the singlet would-be-Goldstone boson mass from that of the pion to
$\eta$ and $\eta'$. Indeed, it has been argued that the pseudoscalar
glueball mass in full QCD is substantially lower than that in the
quenched approximation \cite{GG97}. In view of the fact that the
topological susceptibility is large (of order $10^{-3}\,{\rm GeV}^4$) in
the quenched approximation, and yet is of order $10^{-5}\,{\rm GeV}^4$ in full QCD and zero in the chiral limit, it is conceivable that full QCD has a large
effect on the pseudoscalar glueball as it does on $\eta$ and $\eta'$.

Our conclusion of a lighter pseudoscalar glueball is also supported by a recent analysis based on the chiral Lagrangian with instanton effects \cite{Huang}. Two scenarios for the scalar and pseudoscalar glueball mass difference $\Delta m_G=m(0^{++})-m(0^{-+})$ with the fixed $m(0^{-+})$ were considered in \cite{Huang}. For $0^{++}=f_0(600)$ and $f_0(1710)$, it is found that $\Delta m_G=-(0.1\sim 0.3)$ GeV and $(0.2\sim 1.0)$ GeV, respectively. The first scenario with $0^{++}=f_0(600)$ cannot be realized
since there is no any $0^{-+}$ glueball candidate at energies between $0.7\sim 1.0$ GeV. The second scenario indicates the possible candidates of the $0^{-+}$ glueball are $\eta(1295),~\eta(1405)$ and $\eta(1475)$. The fact that $\eta(1405)$ has a large production rate while $\eta(1295)$ has not been seen in the radiative $J/\psi$ decays and that $\eta(1405)$ has not been observed in $\gamma\gamma$ reactions while $\eta(1475)$ has supports the proposal that $\eta(1405)$ is indeed a good pseudoscalar glueball candidate.
The decay properties of $\eta(1405)$ has been recently studied in \cite{Gutsche,Li}.

\section{Conclusions}
We employed two simple and robust inputs to constrain the mixing matrix of the isoscalar mesons
$f_0(1710)$, $f_0(1500)$, $f_0(1370)$. In the SU(3) symmetry limit, $f_0(1500)$ becomes a pure SU(3) octet and is degenerate with
$a_0(1450)$, while $f_0(1370)$ is mainly an SU(3) singlet with a
slight mixing with the scalar glueball which is the primary
component of $f_0(1710)$. These features remain essentially
unchanged even when SU(3) breaking is taken into account.
From the analysis of the $\eta-\eta'-G$ mixing together
with the inputs from the
recent KLOE experiment, we found a solution for the pseudoscalar
glueball mass around $(1.4\pm 0.1)$ GeV, suggesting that $\eta(1405)$ is indeed a leading candidate for the
pseudoscalar glueball.  Contrary to the mainstream, we thus conjecture that the low-lying pseudoscalar glueball is lighter than the scalar one owing to the dynamic fermion or QCD anomaly effect.

For the lattice community, it will be extremely important to revisit and check the chiral suppression effect in scalar glueball decays. The previous lattice calculation in this respect was done almost 15 yeas ago \cite{Sexton}. If the feature of chiral suppression is confirmed by lattice QCD, it will rule out the possibility of $f_0(1500)$ and $f_0(600)$ as scalar glueball candidates. Although it is a difficult and time-assuming task, a full lattice QCD calculation of the $0^{-+}$ glueball is desperately needed in order to see  if the dynamic fermions will
affect the pseudoscalar glueball spectroscopy dramatically as they do on $\eta$ and $\eta'$.

\begin{theacknowledgments}
I'm very grateful to  Chun-Khiang Chua, Hsiang-nan Li and Keh-Fei Liu  for the fruitful collaboration on glueballs and to the organizers for this great conference.
\end{theacknowledgments}

\end{document}